\documentclass[12pt]{article}
\usepackage{amssymb,amsmath,epsfig}
\allowdisplaybreaks

\begin{document}

\title{\bf Stability Analysis of Einstein Universe in $f(\mathcal{G},T)$ Gravity}
\author{M. Sharif \thanks {msharif.math@pu.edu.pk} and Ayesha Ikram
\thanks{ayeshamaths91@gmail.com}\\
Department of Mathematics, University of the Punjab,\\
Quaid-e-Azam Campus, Lahore-54590, Pakistan.}

\date{}
\maketitle

\begin{abstract}
This paper explores stability of the Einstein universe against
linear homogeneous perturbations in the background of
$f(\mathcal{G},T)$ gravity. We construct static as well as perturbed
field equations and investigate stability regions for the specific
forms of generic function $f(\mathcal{G},T)$ corresponding to
conserved as well as non-conserved energy-momentum tensor. We use
the equation of state parameter to parameterize the stability
regions. The graphical analysis shows that the suitable choice of
parameters lead to stable regions of the Einstein universe.
\end{abstract}

\section{Introduction}

The current accelerated expansion of the universe is one of the most
astonishing discovery in golden era of cosmology. This has
stimulated many researchers to explore the enigmatic nature of dark
energy (DE) which is responsible for the phase of cosmic accelerated
expansion. Dark energy possesses large negative pressure with
repulsive nature but its many salient features are still not known.
Modified theories of gravity are considered as the most favorable
and optimistic approaches among other proposals to explore the
nature of DE. These theories are established by replacing or adding
curvature invariants and their corresponding generic functions in
the geometric part of general relativity (GR).

The Einstein field equations are derived from the first Lovelock
scalar dubbed as the Ricci scalar $(R)$ in the Lagrangian density
which corresponds to gravity while a particular form of quadratic
curvature invariants yields second Lovelock scalar known as
Gauss-Bonnet (GB) invariant. This invariant is a linear combination
of the form
$\mathcal{G}=R^2-4R_{\mu\nu}R^{\mu\nu}+R_{\mu\nu\alpha\beta}
R^{\mu\nu\alpha\beta}$, where $R_{\mu\nu}$ and
$R_{\mu\nu\alpha\beta}$ represent the Ricci and Riemann tensors,
respectively. Gauss-Bonnet invariant is four-dimensional $(4D)$
topological term which has the feature like it is free from spin-2
ghost instabilities \cite{2}. There are two interesting approaches
to discuss the dynamics of $\mathcal{G}$ in $4D$ either by coupling
with scalar field or by adding the generic function $f(\mathcal{G})$
in the Einstein-Hilbert action. The first approach naturally appears
in the effective low energy action in string theory which
effectively discusses the singularity-free cosmological solutions
\cite{3}.

Nojiri and Odintsov \cite{4} introduced second approach as an
alternative for DE known as $f(\mathcal{G})$ gravity which elegantly
studies the fascinating characteristics of late-time cosmology.
Cognola et al. \cite{a} investigated DE cosmology and found that
this theory effectively describes the cosmological structure with a
possibility to describe the transition from decelerated to
accelerated cosmic phases. De Felice and Tsujikawa \cite{5}
constructed some cosmological viable $f(\mathcal{G})$ models and
introduced a procedure to avoid numerical instabilities related with
a large mass of the oscillating mode. The same authors \cite{6} also
found that the solar system constraints are consistent for a wide
range of cosmological viable model parameters.

The captivating issue of cosmic accelerated expansion has
successfully been discussed by taking into account modified theories
of gravity with matter-curvature coupling. Harko et al. \cite{7}
presented $f(R,T)$ gravity ($T$ is the trace of energy-momentum
tensor (EMT)) to study the coupling between geometry and matter.
Recently, we introduced another modified theory named as
$f(\mathcal{G},T)$ gravity which is a generalization of
$f(\mathcal{G})$ gravity \cite{8}. This modification is based on the
coupling of quadratic curvature invariant with matter just as
$f(R,T)$ gravity. We studied the non-zero covariant divergence of
EMT due to matter-curvature coupling and the massive test particles
followed non-geodesic trajectories due to the presence of extra
force while the dust particles moved along geodesic lines of
geometry. In such matter-curvature coupled theories, cosmic
expansion can result from geometric as well as matter component.

The stability issue of the Einstein universe (EU) is as old as
relativistic cosmology. Einstein tried to find static solution of
his field equations to describe isotropic and homogeneous universe.
Since the field equations of GR have no static solution, therefore
Einstein introduced the term known as cosmological constant
$(\Lambda)$ to have static solutions. Einstein universe is described
by static FRW universe model with positive curvature filled with
perfect fluid in the presence of $\Lambda$. Initially, this model is
considered as the most suitable model to discuss static universe but
after few years it is found that EU is unstable against small
isotropic and homogeneous perturbations \cite{10}. Harrison
\cite{11} found that the unstable EU for dust distribution becomes
oscillatory in the presence of radiations and also observed that
stable EU exists against small inhomogeneous perturbations. Gibbons
\cite{12} proved that EU maximizes the entropy against conformal
changes if and only if it is stable against speed of sound $(c_{s})$
greater than $\frac{1}{\sqrt{5}}$. Barrow et al. \cite{13}
demonstrated that EU is always neutrally stable in the presence of
perfect fluid against small inhomogeneous vector as well as tensor
perturbations and also under adiabatic scalar density
inhomogeneities until the inequality $5c_{s}^{2}>1$ holds but
unstable otherwise.

Einstein universe due to its analytical simplicity and fascinating
stability properties has always been of great interest to study in
the extensions of GR as well as in quantum gravity models. Emergent
universe scenario is based on stable EU to resolve the problem of
big-bang singularity which is not successful in GR since EU is
unstable against homogeneous perturbations \cite{b}. To find stable
static solutions, modified theories have gained much attention to
analyze the stability of EU. The stability of EU is studied in
braneworld, Einstein-Cartan theory, loop quantum cosmology,
non-minimal kinetic coupled gravity etc \cite{14}. B\"{o}hmer et al.
\cite{15} explored its stability using scalar homogeneous
perturbations in $f(R)$ gravity and found that stable EU exists for
specific forms of $f(R)$ in contrast to GR. Goswami and his
collaborators \cite{16} investigated the existence as well as
stability of EU in the background of fourth-order gravity theories.
Goheer et al. \cite{17} studied the existence of EU for power-law
$f(R)$ model and found stable solutions. B\"{o}hmer and Lobo
\cite{18} discussed the stability of EU in the context of
$f(\mathcal{G})$ gravity against scalar homogeneous perturbations
and found that stable regions exist for all values of the equation
of state parameter $(\omega)$.

B\"{o}hmer \cite{c} studied the stability of EU parameterized by the
first and second derivatives of scalar potential for linear
homogeneous as well as inhomogeneous perturbations in the context of
hybrid metric-Palatini gravity and found that a large class of
stable solutions. Li et al. \cite{cc} found stable regions for both
open as well as closed universe in modified teleparallel theory
against linear homogeneous scalar perturbations. Huang et al.
\cite{d} obtained stable solutions for EU against homogeneous,
inhomogeneous scalar, tensor and anisotropic perturbations in Jordan
Brans-Dicke theory. The same authors \cite{e} also found the
unstable solutions against homogeneous as well as inhomogeneous
scalar perturbations for open universe while stable EU is obtained
for a closed universe against homogeneous perturbations in
$f(\mathcal{G})$ gravity. B\"{o}hmer and his collaborators \cite{f}
analyzed stability regions against both homogeneous and
inhomogeneous perturbations in scalar-fluid theories and found
stable as well as unstable results against inhomogeneous and
homogeneous perturbations, respectively. Darabi et al. \cite{g}
studied the existence and stability of EU in the context of Lyra
geometry against scalar, vector as well as tensor perturbations for
suitable values of physical parameters. Shabani and Ziaie \cite{19}
analyzed the existence as well as stability of EU in $f(R,T)$
gravity and found stable solutions which were unstable in $f(R)$
gravity.

In this paper, we study the stability of EU against scalar
homogeneous perturbations in the background of $f(\mathcal{G},T)$
gravity. This analysis is helpful to examine the effects of
matter-curvature coupling on the stability of EU. The paper has the
following format. In section \textbf{2}, we construct the field
equations of this theory while section \textbf{3} is devoted to
analyze the stability under linear homogeneous perturbations around
EU for conserved as well as non-conserved EMT. The results are
summarized in the last section.

\section{Dynamics of $f(\mathcal{G},T)$ Gravity}

The action for $f(\mathcal{G},T)$ gravity is given by \cite{8}
\begin{equation}\label{1}
\mathcal{S}=\int
d^{4}x\sqrt{-g}\left[\frac{R+f(\mathcal{G},T)}{2\kappa^2}
+\mathcal{L}_{m}\right],
\end{equation}
where $T=g_{\mu\nu}T^{\mu\nu},~\kappa^2,~g$ and $\mathcal{L}_{m}$
represent coupling constant, determinant of the metric tensor
$(g_{\mu\nu})$ and matter Lagrangian density, respectively. The EMT
in terms of $\mathcal{L}_{m}$ is defined as \cite{20}
\begin{equation}\label{2}
T_{\mu\nu}=-\frac{2}{\sqrt{-g}}
\frac{\delta(\sqrt{-g}\mathcal{L}_{m})}{\delta g^{\mu\nu}}.
\end{equation}
If $\mathcal{L}_{m}$ depends on the components of $g_{\mu\nu}$ but
does not depend on its derivatives, then Eq.(\ref{2}) yields
\begin{equation}\label{3}
T_{\mu\nu}=g_{\mu\nu}\mathcal{L}_{m}-2\frac{\partial
\mathcal{L}_{m}}{\partial g^{\mu\nu}}.
\end{equation}
Varying the action (\ref{1}) with respect to $g_{\mu\nu}$, we obtain
the field equations as follows
\begin{eqnarray}\nonumber
R_{\mu\nu}-\frac{1}{2}g_{\mu\nu}R&=&\kappa^2T_{\mu\nu}
-(T_{\mu\nu}+\Theta_{\mu\nu})f_{T}(\mathcal{G},T)+\frac{1}{2}g_{\mu\nu}
f(\mathcal{G},T)-(2RR_{\mu\nu}\\\nonumber&-&4R_{\mu}^{\alpha}R_{\alpha\nu}
-4R_{\mu\alpha\nu\beta}R^{\alpha\beta}+2R_{\mu}^{\alpha\beta\gamma}
R_{\nu\alpha\beta\gamma})f_{\mathcal{G}}(\mathcal{G},T)
\\\nonumber&-&(2Rg_{\mu\nu}
\Box-4R_{\mu\nu}\Box-2R\nabla_{\mu}\nabla_{\nu}
+4R_{\mu}^{\alpha}\nabla_{\nu}\nabla_{\alpha}
+4R_{\nu}^{\alpha}\nabla_{\mu}\nabla_{\alpha}\\\label{4}&-&4g_{\mu\nu}R^{\alpha\beta}
\nabla_{\alpha}\nabla_{\beta}+4R_{\mu\alpha\nu\beta}
\nabla^{\alpha}\nabla^{\beta})f_{\mathcal{G}}(\mathcal{G},T),
\end{eqnarray}
where $f_{\mathcal{G}}(\mathcal{G},T)=\partial
f(\mathcal{G},T)/\partial\mathcal{G},~f_{T}(\mathcal{G},T)=\partial
f(\mathcal{G},T)/\partial T,~\Box=\nabla_{\mu}\nabla^{\mu}$ and
$\nabla_{\mu}$ is a covariant derivative whereas $\Theta_{\mu\nu}$
has the following expression
\begin{equation}\label{5}
\Theta_{\mu\nu}=g^{\alpha\beta}\frac{\delta T_{\alpha\beta}}{\delta
g_{\mu\nu}}=-2T_{\mu\nu}+g_{\mu\nu}\mathcal{L}_{m}-2g^{\alpha\beta}
\frac{\partial^{2}\mathcal{L}_{m}}{\partial g^{\mu\nu}\partial
g^{\alpha\beta}}.
\end{equation}
The covariant divergence of Eq.(\ref{4}) is given by
\begin{eqnarray}\nonumber
\nabla^{\mu}T_{\mu\nu}&=&\frac{f_{T}(\mathcal{G},T)}
{\kappa^2-f_{T}(\mathcal{G},T)}\left[\nabla^{\mu}\Theta_{\mu\nu}
-\frac{1}{2}g_{\mu\nu}\nabla^{\mu}T+(T_{\mu\nu}+\Theta_{\mu\nu})
\right.\\\label{6}&\times&\left.\nabla^{\mu}(\ln{f_{T}(\mathcal{G},T)})
\right].
\end{eqnarray}

In this theory, the field equations as well as conservation law
depend on the contributions from cosmic matter contents, therefore
every suitable selection of $\mathcal{L}_{m}$ provides the
particular scheme of dynamical equations. The line element for
positive curvature FRW universe model is \cite{15}
\begin{equation}\label{7}
ds^2=dt^2-a^2(t)\left(\frac{1}{1-r^2}dr^2+r^2(d\theta^2+\sin^2\theta
d\phi^2)\right),
\end{equation}
where $a(t)$ is the scale factor. The energy-momentum tensor for
perfect fluid is given by
\begin{equation}\label{8}
T_{\mu\nu}=(\rho+P)u_{\mu}u_{\nu}-Pg_{\mu\nu},
\end{equation}
where $\rho,~P$ and $u_{\mu}$ represent the energy density, pressure
and four-velocity of the matter distribution, respectively. For
perfect fluid as cosmic matter distribution with
$\mathcal{L}_{m}=-P$, Eq.(\ref{5}) becomes \cite{7}
\begin{equation}\label{9}
\Theta_{\mu\nu}=-2T_{\mu\nu}-Pg_{\mu\nu}.
\end{equation}
Using Eqs.(\ref{7})-(\ref{9}) in (\ref{4}), we obtain the following
set of field equations
\begin{eqnarray}\nonumber
\frac{3}{a^{2}}(1+\dot{a}^2)&=&\kappa^2\rho+\frac{1}{2}f(\mathcal{G},T)
+(\rho+P)f_{T}(\mathcal{G},T)-12\frac{\ddot{a}}{a^3}(1+\dot{a}^2)
\\\label{10}&\times&f_{\mathcal{G}}(\mathcal{G},T)+12\frac{\dot{a}}{a^3}
(1+\dot{a}^2)\partial_{t}f_{\mathcal{G}}(\mathcal{G},T),\\\nonumber
-(1+\dot{a}^2)-2a\ddot{a}&=&\kappa^2a^2P-\frac{1}{2}a^2f(\mathcal{G},T)
+12\frac{\ddot{a}}{a}(1+\dot{a}^2)f_{\mathcal{G}}(\mathcal{G},T)
\\\label{11}&-&8\dot{a}\ddot{a}\partial_{t}f_{\mathcal{G}}(\mathcal{G},T)
-4(1+\dot{a}^2)\partial_{tt}f_{\mathcal{G}}(\mathcal{G},T),
\end{eqnarray}
where
\begin{equation}\label{12}
\mathcal{G}=\frac{24}{a^{3}}(1+\dot{a}^{2})\ddot{a},\quad T=\rho-3P,
\end{equation}
and dot represents the time derivative. The conservation equation
(\ref{6}) for perfect fluid yields
\begin{eqnarray}\nonumber
\dot{\rho}+3\frac{\dot{a}}{a}(\rho+P)&=&\frac{-1}{\kappa^2
+f_{T}(\mathcal{G},T)}\left[\left(\dot{P}+\frac{1}{2}\dot{T}\right)
f_{T}(\mathcal{G},T)+(\rho+P)\right.\\\label{12a}&\times&
\left.\partial_{t}f_{T}(\mathcal{G},T)\right].
\end{eqnarray}

\section{Stability of Einstein Universe}

In this section, we analyze the stability of EU against linear
homogeneous perturbations in the background of $f(\mathcal{G},T)$
gravity. For EU, $a(t)=a_{0}=$ constant and consequently, the field
equations (\ref{10}) and (\ref{11}) reduce to
\begin{eqnarray}\label{13}
\frac{3}{a_{0}^2}&=&\kappa^2\rho_{0}+\frac{1}{2}
f(\mathcal{G}_{0},T_{0})+(\rho_{0}+P_{0})f_{T}(\mathcal{G}_{0},T_{0}),
\\\label{14}-\frac{1}{a_{0}^2}&=&\kappa^2P_{0}-\frac{1}{2}f(\mathcal{G}_{0},T_{0}),
\end{eqnarray}
where
$\mathcal{G}_{0}=\mathcal{G}(a_{0})=0,~T_{0}=\rho_{0}-3P_{0},~\rho_{0}$
and $P_{0}$ are the unperturbed energy density and pressure,
respectively. To explore the stability regions, we consider linear
form of equation of state as $P(t)=\omega\rho(t)$ and define linear
perturbations in the scale factor and energy density as follows
\begin{equation}\label{15}
a(t)=a_{0}+a_{0}\delta
a(t),\quad\rho(t)=\rho_{0}+\rho_{0}\delta\rho(t),
\end{equation}
where $\delta a(t)$ and $\delta \rho(t)$ represent the perturbed
scale factor and energy density, respectively. Applying the Taylor
series expansion in two variables upto first order with the
assumption that $f(\mathcal{G},T)$ is an analytic function, we have
\begin{equation}\label{16}
f(\mathcal{G},T)=f(\mathcal{G}_{0},T_{0})+
f_{\mathcal{G}}(\mathcal{G}_{0},T_{0})\delta\mathcal{G}
+f_{T}(\mathcal{G}_{0},T_{0})\delta T,
\end{equation}
where $\delta\mathcal{G}$ and $\delta T$ have the following
expressions
\begin{equation}\label{17}
\delta\mathcal{G}=\frac{24}{a_{0}^{2}}\delta\ddot{a},\quad \delta
T=T_{0}\delta\rho,
\end{equation}
where $\delta\ddot{a}=\frac{d^2}{dt^2}(\delta a)$. Using
Eqs.(\ref{13})-(\ref{17}) in (\ref{10}) and (\ref{11}), we obtain
the linearized perturbed field equations as follows
\begin{eqnarray}\nonumber
&&6\delta a+24\rho_{0}(1+\omega)
f_{\mathcal{G}T}(\mathcal{G}_{0},T_{0})\delta\ddot{a}+a_{0}^2\rho_{0}[\kappa^2+(1+\omega)
f_{T}(\mathcal{G}_{0},T_{0})\\\label{18}&+&\frac{1}{2}(1-3\omega)
f_{T}(\mathcal{G}_{0},T_{0})+\rho_{0}(1+\omega)(1-3\omega)
f_{TT}(\mathcal{G}_{0},T_{0})]\delta\rho=0,\\\nonumber
&-&\frac{2}{a_{0}^2}\delta a+2\delta\ddot{a}-\frac{96}{a_{0}^4}
f_{\mathcal{GG}}(\mathcal{G}_{0},T_{0})\delta a^{(iv)}+\rho_{0}
[\kappa^2\omega-\frac{1}{2}(1-3\omega)f_{T}(\mathcal{G}_{0},T_{0})]
\delta\rho\\\label{19}&-&4\frac{\rho_{0}}{a_{0}^2}(1-3\omega)
f_{\mathcal{G}T}(\mathcal{G}_{0},T_{0})\delta\ddot{\rho}=0.
\end{eqnarray}
These equations show that the perturbations in $a(t)$ are related
with density perturbations. In the following subsections, we discuss
the stability modes for conserved as well as non-conserved EMT.

\subsection{Conserved EMT}

In this case, we assume that general conservation law holds in
$f(\mathcal{G},T)$ gravity. For this purpose, the right hand side of
Eq.(\ref{12a}) must be zero which yields
\begin{equation}\label{21}
\left(\dot{P}+\frac{1}{2}\dot{T}\right)f_{T}(\mathcal{G},T)+
(\rho+P)\partial_{t}f_{T}(\mathcal{G},T)=0.
\end{equation}
The conserved matter contents of the universe satisfy the relation
given by
\begin{equation}\label{22}
\delta\dot{\rho}=-3(1+\omega)\delta\dot{a}.
\end{equation}
Using this equation in the elimination of $\delta\rho$ from
Eqs.(\ref{18}) and (\ref{19}), we obtain the fourth-order
perturbation equation in perturbed $a(t)$ as follows
\begin{eqnarray}\nonumber
&&\left[6\kappa^2a_{0}\omega-3a_{0}(1-3\omega)f_{T}
(\mathcal{G}_{0},T_{0})+2a_{0}\left\{\kappa^2+(1+\omega)f_{T}
(\mathcal{G}_{0},T_{0})\right.\right.\\\nonumber&+&\left.
\left.\frac{1}{2}(1-3\omega)f_{T}(\mathcal{G}_{0},T_{0})+\rho_{0}(1+\omega)(1-3\omega)
f_{TT}(\mathcal{G}_{0},T_{0})\right\}\right]\delta
a\\\nonumber&+&\left[24a_{0}\rho_{0}(1+\omega)\left\{
\kappa^2\omega-\frac{1}{2}(1-3\omega)f_{T}(\mathcal{G}_{0},T_{0})
\right\}f_{\mathcal{G}T}(\mathcal{G}_{0},T_{0})\right.\\\nonumber&-&\left.
\left\{2+\frac{12\rho_{0}}{a_{0}^2}(1+\omega)(1-3\omega)f_{\mathcal{G}T}
(\mathcal{G}_{0},T_{0})\right\}\left\{\kappa^2a_{0}^3+a_{0}^3(1+\omega)
f_{T}(\mathcal{G}_{0},T_{0})\right.\right.\\\nonumber&+&\left.\left.
\frac{1}{2}a_{0}^3(1-3\omega)f_{T}(\mathcal{G}_{0},T_{0})+a_{0}^3
\rho_{0}(1+\omega)(1-3\omega)f_{TT}(\mathcal{G}_{0},T_{0})\right\}\right]
\delta\ddot{a}\\\nonumber&+&\frac{96}{a_{0}^4}\left\{\kappa^2a_{0}^3
+a_{0}^3(1+\omega)f_{T}(\mathcal{G}_{0},T_{0})+\frac{1}{2}a_{0}^3(1-3\omega)
f_{T}(\mathcal{G}_{0},T_{0})+a_{0}^3\rho_{0}\right.\\\label{23}&\times&
\left.(1+\omega)(1-3\omega)f_{TT}(\mathcal{G}_{0},T_{0})\right\}
f_{\mathcal{GG}}(\mathcal{G}_{0},T_{0})\delta a^{(iv)}=0.
\end{eqnarray}
Adding Eqs.(\ref{13}) and (\ref{14}), it follows that
\begin{equation}\label{24}
\frac{2}{a_{0}^2}=\rho_{0}(1+\omega)(\kappa^2+f_{T}(\mathcal{G}_{0},T_{0})).
\end{equation}
Using this expression in Eq.(\ref{23}), the resulting perturbation
equation yields
\begin{eqnarray}\nonumber
&&\left[\rho_{0}(1+\omega)\{\kappa^2+f_{T}(\mathcal{G}_{0},T_{0})\}\left\{
\kappa^2(1+3\omega)+(1+\omega)f_{T}(\mathcal{G}_{0},T_{0})
\right.\right.\\\nonumber&-&\left.\left.(1-3\omega)
f_{T}(\mathcal{G}_{0},T_{0})+\rho_{0}(1+\omega)(1-3\omega)f_{TT}
(\mathcal{G}_{0},T_{0})\right\}\right]\delta
a\\\nonumber&+&\left[12\rho_{0}^2(1+\omega)^2\{\kappa^2+f_{T}
(\mathcal{G}_{0},T_{0})\}\left\{\kappa^2\omega-\frac{1}{2}
(1-3\omega)f_{T}(\mathcal{G}_{0},T_{0})\right\}\right.
\\\nonumber&\times&\left.f_{\mathcal{G}T}(\mathcal{G}_{0},T_{0})
-[2+6\rho_{0}^2(1+\omega)^2(1-3\omega)\{\kappa^2+f_{T}
(\mathcal{G}_{0},T_{0})\}f_{\mathcal{G}T}(\mathcal{G}_{0},T_{0})]
\right.\\\nonumber&\times&\left.\left\{\kappa^2+(1+\omega)
f_{T}(\mathcal{G}_{0},T_{0})+\rho_{0}(1+\omega)(1-3\omega)
f_{TT}(\mathcal{G}_{0},T_{0})+\frac{1}{2}(1-3\omega)\right.\right.
\\\nonumber&\times&\left.\left.f_{T}(\mathcal{G}_{0},T_{0})\right\}
\right]\delta\ddot{a}+24\rho_{0}^2(1+\omega)^2\{\kappa^2+f_{T}
(\mathcal{G}_{0},T_{0})\}^2\left\{\kappa^2+(1+\omega)\right.
\\\nonumber&\times&\left.f_{T}(\mathcal{G}_{0},T_{0})
+\rho_{0}(1+\omega)(1-3\omega)f_{TT}(\mathcal{G}_{0},T_{0})+\frac{1}{2}
(1-3\omega)f_{T}(\mathcal{G}_{0},T_{0})\right\}\\\label{25}&\times&
f_{\mathcal{GG}}(\mathcal{G}_{0},T_{0})\delta a^{(iv)}=0.
\end{eqnarray}

The solution of this equation helps to discuss the stability regions
in EU. However, it would be difficult to find stable/unstable
solutions due to its complicated nature. We, therefore, consider the
particular form of $f(\mathcal{G},T)$ as follows
\begin{equation}\label{26}
f(\mathcal{G},T)=f_{1}(\mathcal{G})+f_{2}(T).
\end{equation}
This choice of model does not involve the direct curvature-matter
non-minimal coupling but it can be considered as correction to
$f(\mathcal{G})$ gravity. In this case, we have assumed that the EMT
is conserved, therefore, we first constrain the above model such
that the conservation law holds for it. For this purpose, using the
considered form in Eq.(\ref{21}), the resulting second order
differential equation takes the form
\begin{equation}\nonumber
(1-\omega)f_{2}'(T)+2(1+\omega)Tf_{2}''(T)=0,
\end{equation}
where prime represents derivative with respect to $x~(x=T$ or
$\mathcal{G})$. The solution is given by
\begin{equation}\label{27}
f_{2}(T)=\frac{c_{1}(1+\omega)}{1+3\omega}T^{\frac{1+3\omega}{2(1+\omega)}}+c_{2},
\end{equation}
where $c_{i}$'s $(i=1,2)$ are integration constants. This is the
unique representation of matter contribution for which conservation
law holds with model (\ref{26}). The modified GB term
$f_{1}(\mathcal{G}_{0})$ acts like an effective $\Lambda$ to the
unperturbed field equations. It is worth mentioning here that
$f(\mathcal{G})$ gravity is recovered for this choice of
$f(\mathcal{G},T)$ model if $f_{2}(T)=0$ \cite{18}. Inserting the
values from Eqs.(\ref{26}) and (\ref{27}) in (\ref{25}), the
differential equation takes the form
\begin{equation}\label{28}
\Delta_{2}(\Delta_{1}+\Delta_{3})\delta a-2\Delta_{1}\delta
\ddot{a}+24\Delta_{1}\Delta_{2}^{2}f_{1}''(\mathcal{G}_{0})\delta
a^{(iv)}=0,
\end{equation}
where $\Delta_{j}$'s $(j=1,2,3)$ are
\begin{eqnarray}\nonumber
\Delta_{1}&=&\kappa^2+\frac{1}{4}c_{1}(1+5\omega)[\rho_{0}(1-3
\omega)]^{\frac{\omega-1}{2(\omega+1)}}-\frac{1}{4}c_{1}\rho_{0}
(1-\omega)(1-3\omega)\\\nonumber&\times&[\rho_{0}(1-3\omega)]
^{\frac{-(3+\omega)}{2(1+\omega)}},\\\nonumber\Delta_{2}&=&
\rho_{0}(1+\omega)\left[\kappa^2+\frac{1}{2}\{\rho_{0}(1-3\omega)\}
^{\frac{\omega-1}{2(\omega+1)}}\right],\\\nonumber\Delta_{3}&=&
3\kappa^2\omega-\frac{3}{4}c_{1}(1-3\omega)\{\rho_{0}(1-3\omega)\}
^{\frac{\omega-1}{2(\omega+1)}}.
\end{eqnarray}
Equation (\ref{28}) provides the following solution
\begin{equation}\nonumber
\delta
a(t)=d_{1}e^{\Omega_{1}t}+d_{2}e^{-\Omega_{1}t}+d_{3}e^{\Omega_{2}t}
+d_{4}e^{-\Omega_{2}t},
\end{equation}
where $d_{k}$'s $(k=1...4)$ are constants of integration and the
parameters $\Omega_{1}$ and $\Omega_{2}$ are frequencies of small
perturbations given by
\begin{equation}\label{29}
\Omega^{2}_{1,2}=\frac{\Delta_{1}\pm\sqrt{\Delta_{1}^{2}-24
\Delta_{1}\Delta_{2}^{3}(\Delta_{1}+\Delta_{3})f_{1}''
(\mathcal{G}_{0})}}{24\Delta_{1}\Delta_{2}^{2}f_{1}''
(\mathcal{G}_{0})}.
\end{equation}

In order to avoid the exponential growth of $\delta a(t)$ or
collapse, the frequencies are purely complex which lead to the
existence of stable EU. Thus, the condition of stability is achieved
when $\Omega^{2}_{1,2}<0$. In the limit of GR, $\Omega^{2}_{1}$
diverge while $\Omega^{2}_{2}$ are given by
\begin{equation}\nonumber
\Omega^{2}_{2}=\frac{1}{2}\kappa^2\rho_{0}(1+3\omega)(1+\omega),
\end{equation}
which provide stable region in the range  $-1<\omega<-\frac{1}{3}$
\cite{18}. For simplicity, we introduce a new parameter
$\zeta_{1}=24f_{1}''(\mathcal{G}_{0})$ as well as use $\kappa^2=1$
and $\rho_{0}=0.3$ (present day value of density parameter) to
discuss the graphical analysis of stable EU \cite{h2}. Figure
\textbf{1} shows the stable regions under homogeneous perturbations
of EU for $\Omega_{1}^{2}$. It is found that for $c_{1}=1$ in the
left plot, the stable EU exists for negative values of $\omega$
while no stable region exists for its positive values. The right
panel shows the stable region for $c_{1}=5$ and hence the stability
regions decrease as the value of integration constant increases
while for negative values of $c_{1}$, no stable regions are found.
The regions of stability for frequencies $\Omega_{2}^{2}$ are shown
in Figure \textbf{2} for both positive as well as negative values of
$c_{1}$. The negative values of $\zeta_{1}$ are obtained for
$f_{1}''(\mathcal{G}_{0})<0$ which is in agreement with stability
condition of $f(\mathcal{G})$ models \cite{21}. Figure \textbf{3}
shows the stability regions for both $\Omega_{1}^{2}$ as well as
$\Omega_{2}^{2}$ of the whole system.
\begin{figure}
\epsfig{file=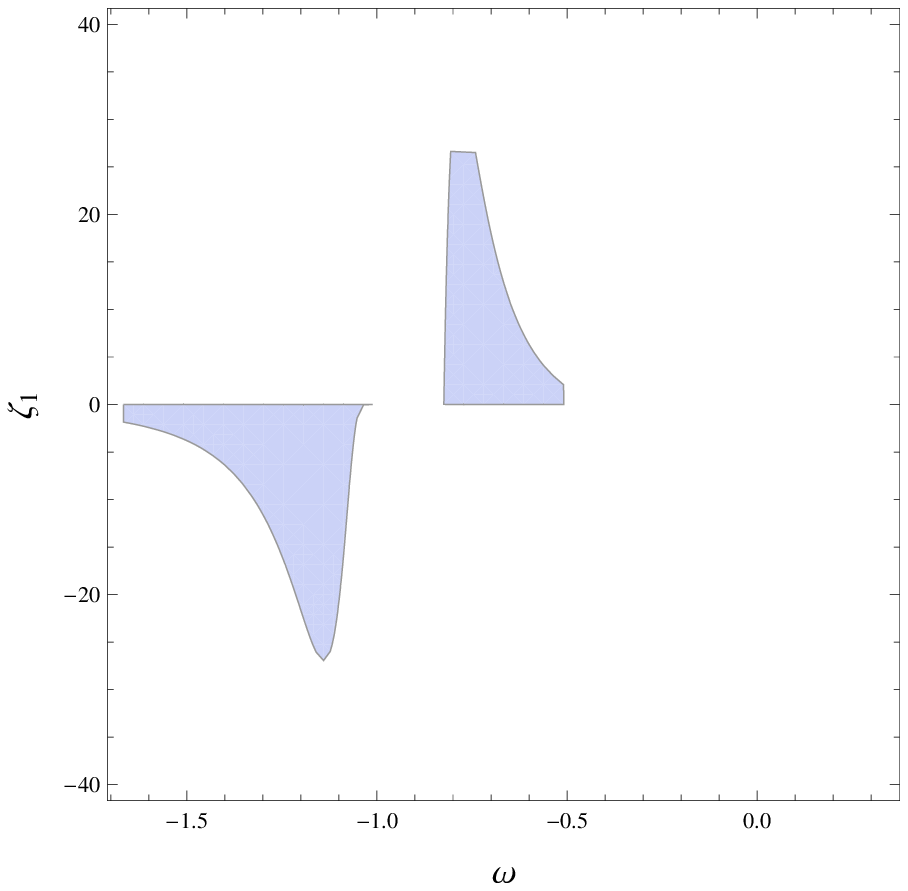, width=0.5\linewidth}\epsfig{file=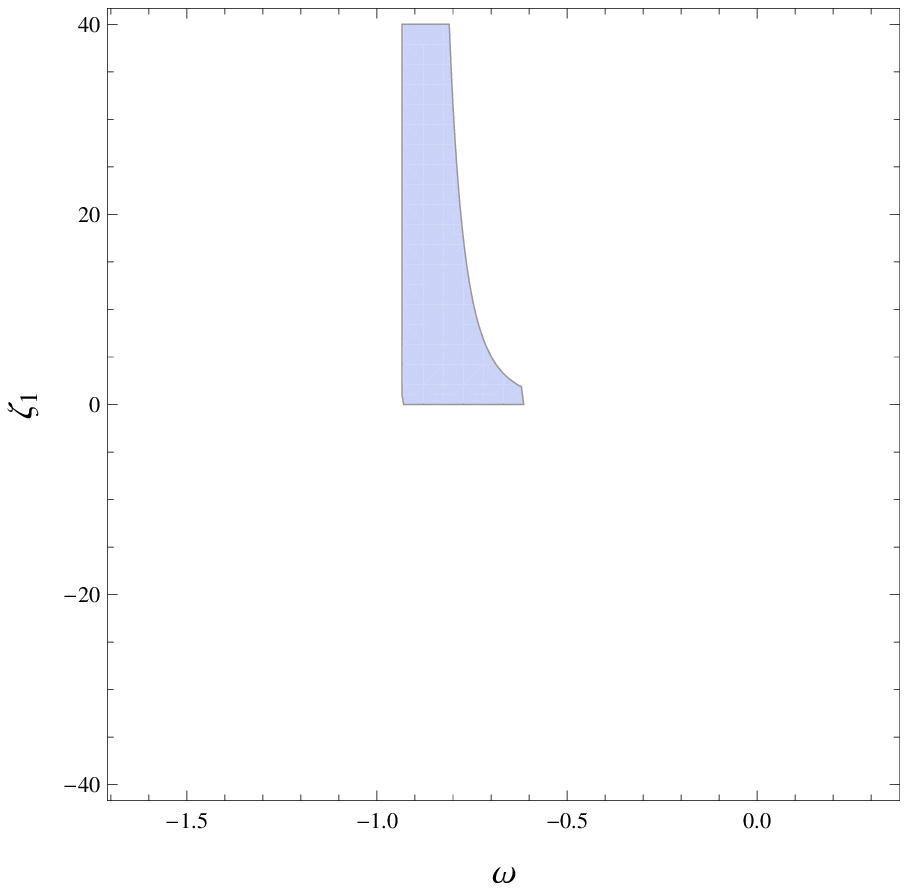,
width=0.5\linewidth}\caption{Stable regions in $(\omega,\zeta_{1})$
space for $\Omega_{1}^{2}$ with $c_{1}=1$ (left) and $c_{1}=5$
(right).}
\end{figure}
\begin{figure}
\epsfig{file=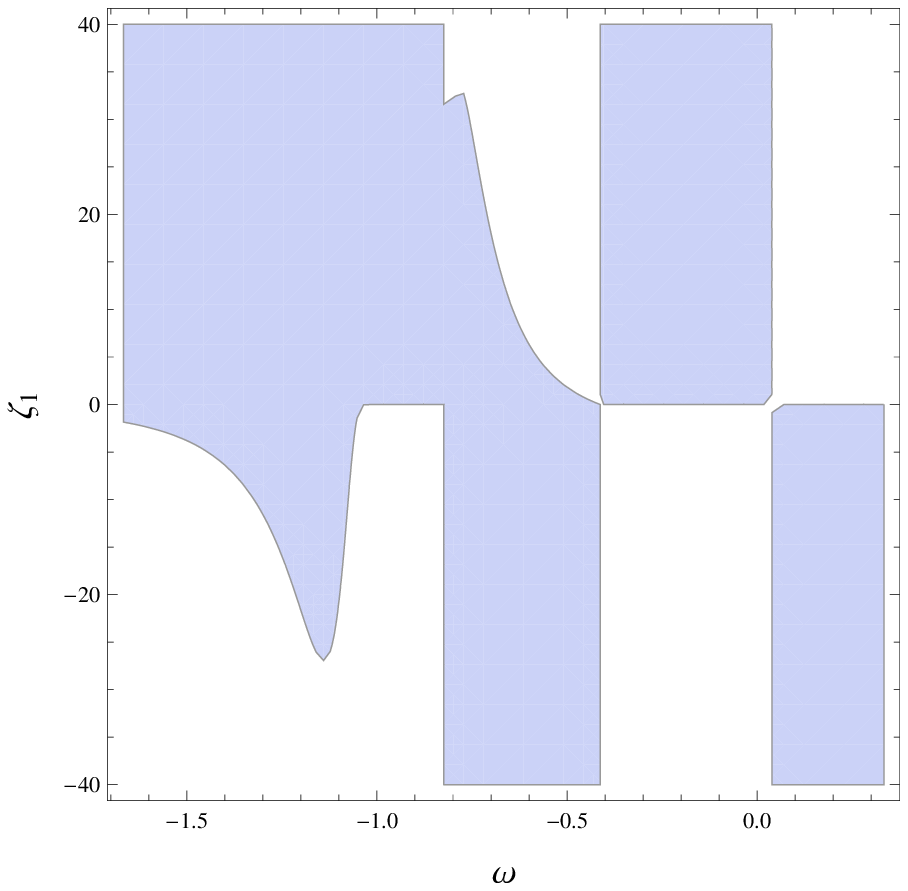, width=0.5\linewidth}\epsfig{file=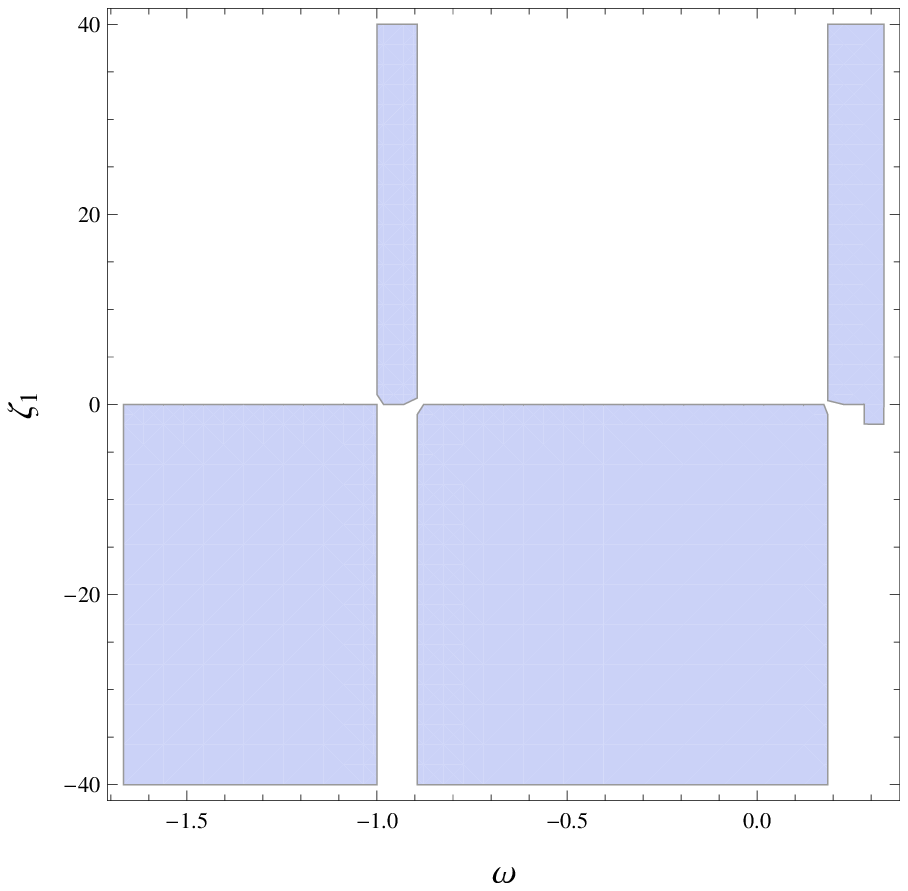,
width=0.5\linewidth}\caption{Stable regions in $(\omega,\zeta_{1})$
space for $\Omega_{2}^{2}$ with $c_{1}=1$ (left) and $c_{1}=-1$
(right).}
\end{figure}
\begin{figure}
\epsfig{file=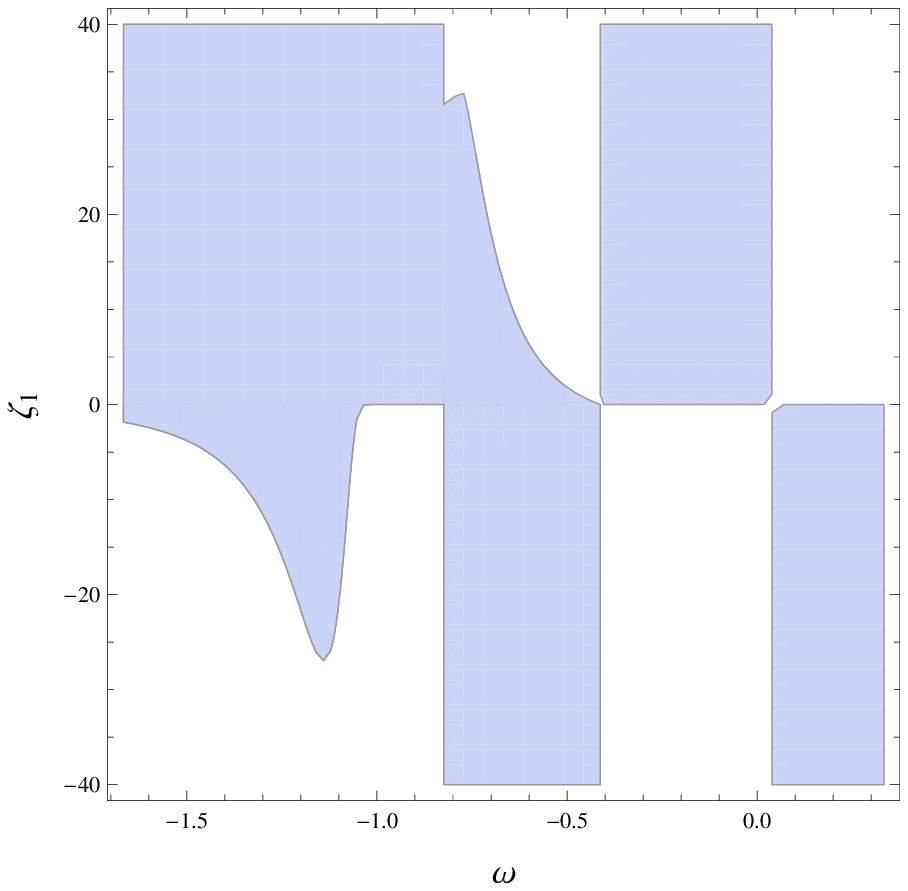, width=0.5\linewidth}\epsfig{file=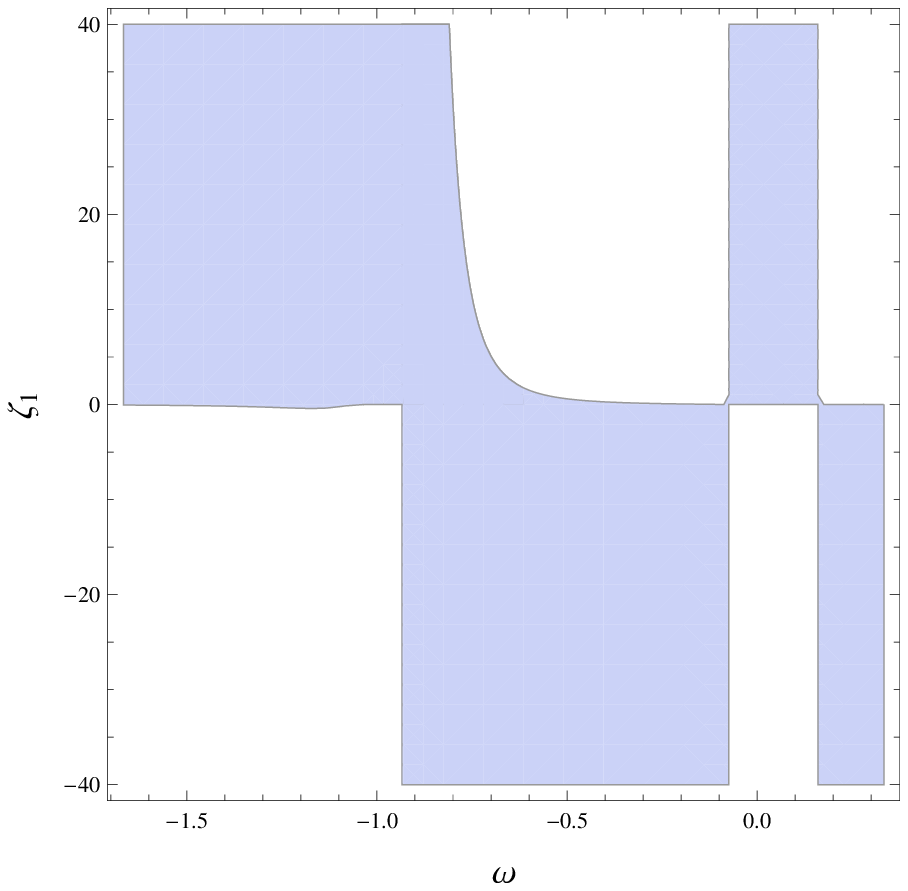,
width=0.5\linewidth}\caption{Stable regions in $(\omega,\zeta_{1})$
space for $\Omega_{1,2}^{2}$ with $c_{1}=1$ (left) and $c_{1}=5$
(right).}
\end{figure}

\subsection{Non-Conserved EMT}

Here, we analyze the stability of $f(\mathcal{G},T)$ model when EMT
is not conserved. We consider generic function $f_{1}(\mathcal{G})$
and a linear form of $f_{2}(T)$ in Eq.(\ref{26}) as follows
\begin{equation}\label{30}
f(\mathcal{G},T)=f_{1}(\mathcal{G})+\kappa^{2}\chi T,
\end{equation}
where $\chi$ is an arbitrary constant. Substituting in
Eq.(\ref{12a}), we obtain
\begin{equation}\nonumber
\rho=\tilde{\rho}_{0}a^{-3\varphi},\quad\varphi=\frac{2(1+\chi)(1+\omega)}
{2+\chi(3-\omega)},
\end{equation}
where $\tilde{\rho}_{0}$ is an integration constant. The perturbed
field equations (\ref{18}) and (\ref{19}) take the following form
\begin{eqnarray}\label{31}
&&6\delta
a+\kappa^2a_{0}^2\rho_{0}\left[1-\frac{\chi}{2}(\omega-3)\right]\delta\rho=0,
\\\label{32}&&2\delta\ddot{a}-\frac{2}{a_{0}^2}\delta
a+\kappa^2\left[\omega-\frac{\chi}{2}(1-3\omega)\right]\rho_{0}\delta\rho
-\frac{96}{a_{0}^{4}}f_{1}''(\mathcal{G}_{0})\delta a^{(iv)}=0.
\end{eqnarray}
The first field equation shows the relationship between the
perturbed energy density and scale factor perturbations. Eliminating
$\delta\rho$ from Eqs.(\ref{31}) and (\ref{32}), the resulting
differential equation in perturbed $a(t)$ is given by
\begin{equation}\label{33}
2\delta\ddot{a}-\frac{2}{a_{0}^2}\left[1+\frac{3(2\omega-\chi(1-3\omega))}
{2-\chi(\omega-3)}\right]\delta
a-\frac{96}{a_{0}^4}f_{1}''(\mathcal{G}_{0})\delta a^{(iv)}=0.
\end{equation}
In this case, the addition of static field equations yields
\begin{equation}\label{34}
\frac{2}{a_{0}^2}=\kappa^2\rho_{0}(1+\chi)(1+\omega).
\end{equation}
Inserting this value of $\frac{2}{a_{0}^2}$ in Eq.(\ref{33}), we
obtain
\begin{eqnarray}\nonumber
&&\kappa^2\rho_{0}[\chi(1+\chi)(1-\omega^2)-(1+\chi)^2(1+\omega)(1+3\omega)]\delta
a+[2(1+\chi)\\\nonumber&+&\chi(1-\omega)]\delta\ddot{a}-12\kappa^4
\rho_{0}^{2}[2(1+\chi)^{3}(1+\omega)^{2}+\chi(1+\chi)^{2}(1-\omega)(1+\omega)^{2}]
\\\label{35}&\times&f_{1}''(\mathcal{G}_{0})\delta a^{(iv)}=0,
\end{eqnarray}
whose solution provides the following four frequencies as
\begin{equation}\nonumber
\Upsilon^{2}_{1,2}=\frac{-2(1+\chi)-\chi(1-\omega)\pm\sqrt{[2(1+\chi)
+\chi(1-\omega)]^{2}-48\kappa^{6}\rho_{0}^{3}\Delta_{4}f_{1}''(\mathcal{G}_{0})}}{24
\kappa^4\rho_{0}^{2}[\chi(1+\chi)^{2}(\omega-1)(1+\omega)^{2}
-2(1+\chi)^{3}(1+\omega)^{2}]f_{1}''(\mathcal{G}_{0})},
\end{equation}
where
\begin{eqnarray}\nonumber
\Delta_{4}&=&[2(1+\chi)^{3}(1+\omega)^{2}+\chi(1+\chi)^{2}(1-\omega)(1+\omega)^{2}]
[(1+\chi)^2(1+\omega)\\\nonumber&\times&(1+3\omega)-\chi(1+\chi)(1-\omega^2)].
\end{eqnarray}
\begin{figure}
\epsfig{file=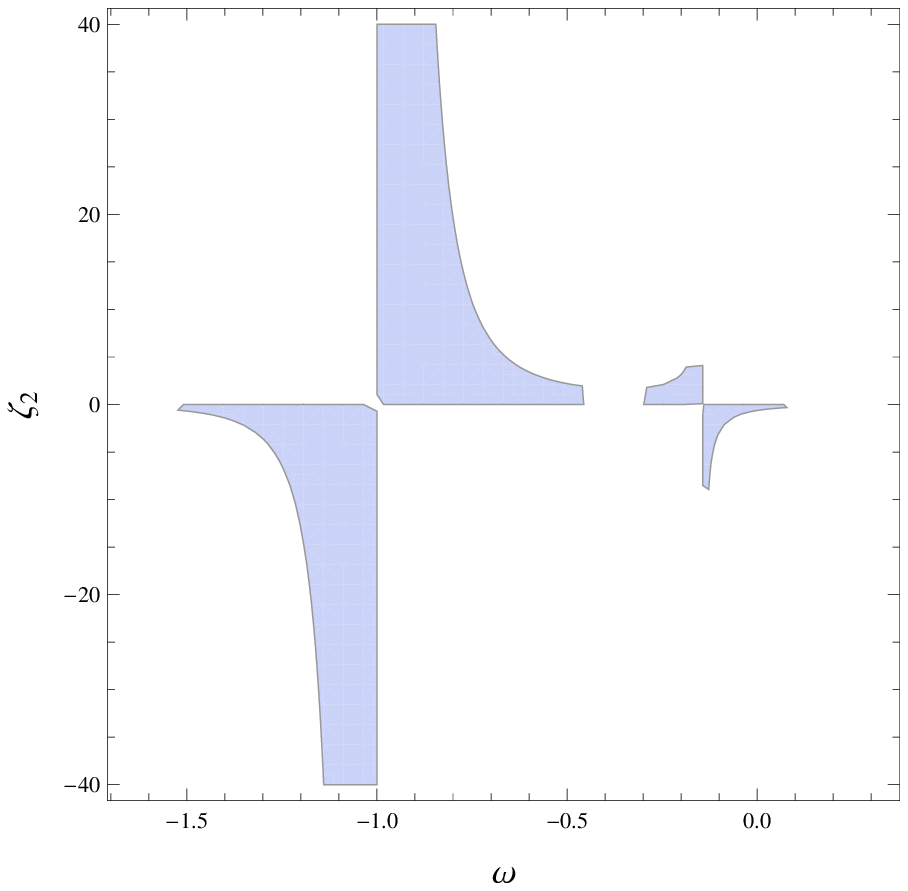, width=0.5\linewidth}\epsfig{file=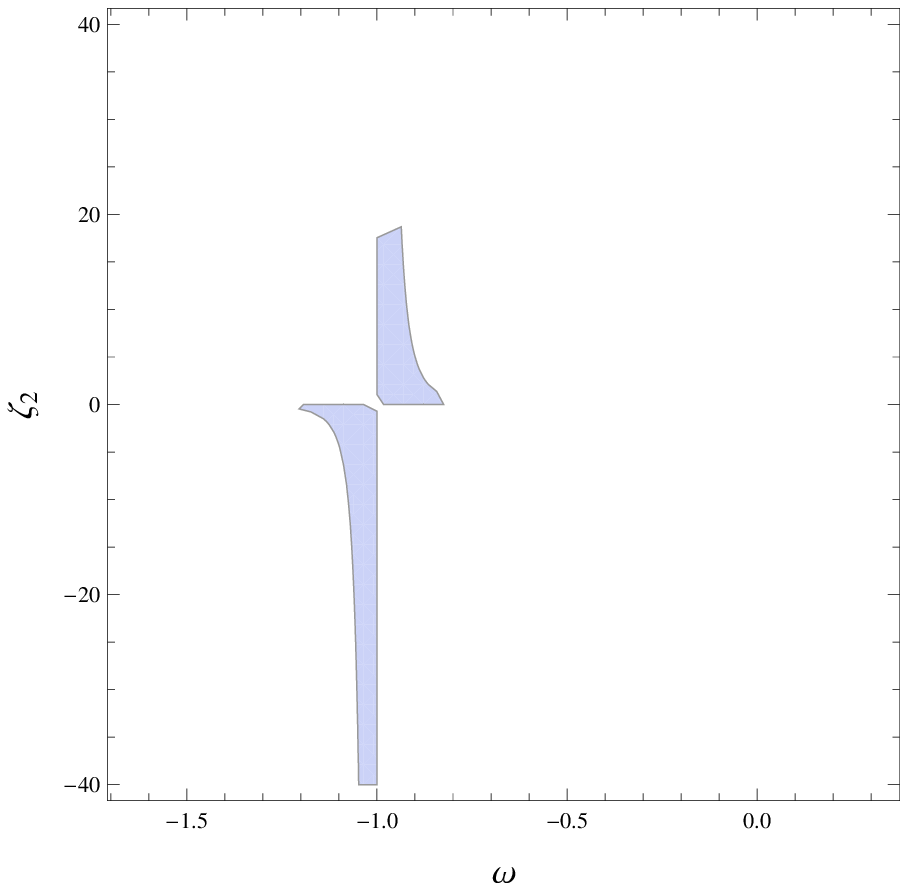,
width=0.5\linewidth}\caption{Stable regions in $(\omega,\zeta_{2})$
space for $\Upsilon_{1}^{2}$ with $\chi=1$ (left) and $\chi=5$
(right).}
\end{figure}
\begin{figure}
\epsfig{file=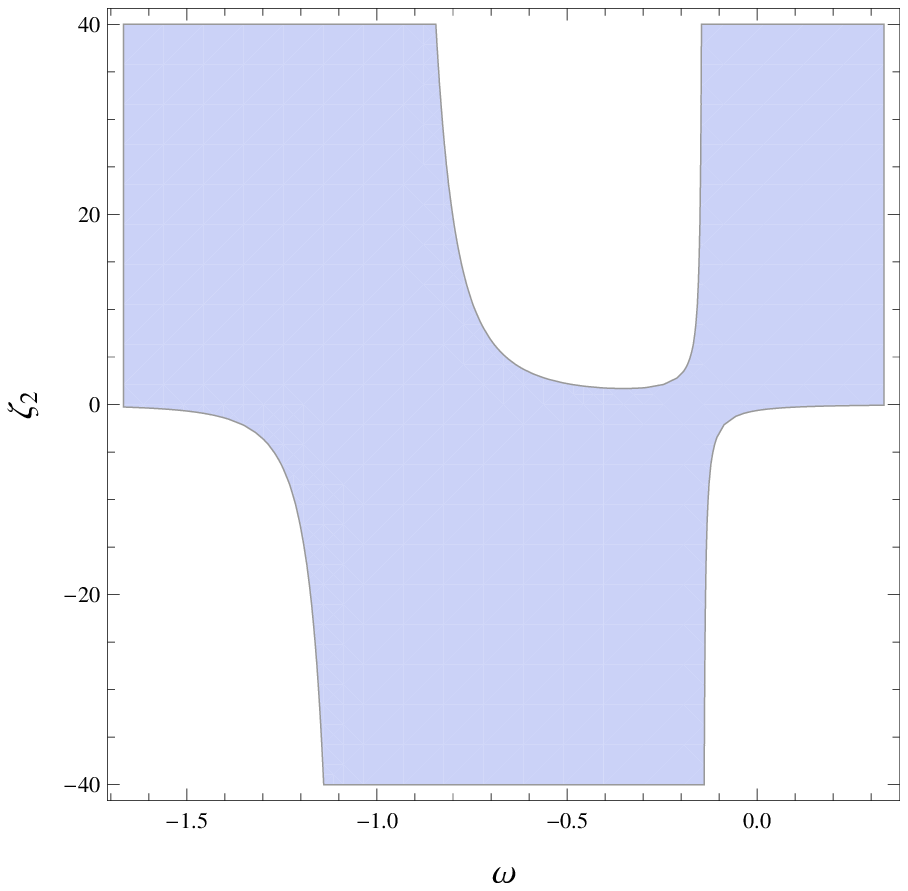, width=0.5\linewidth}\epsfig{file=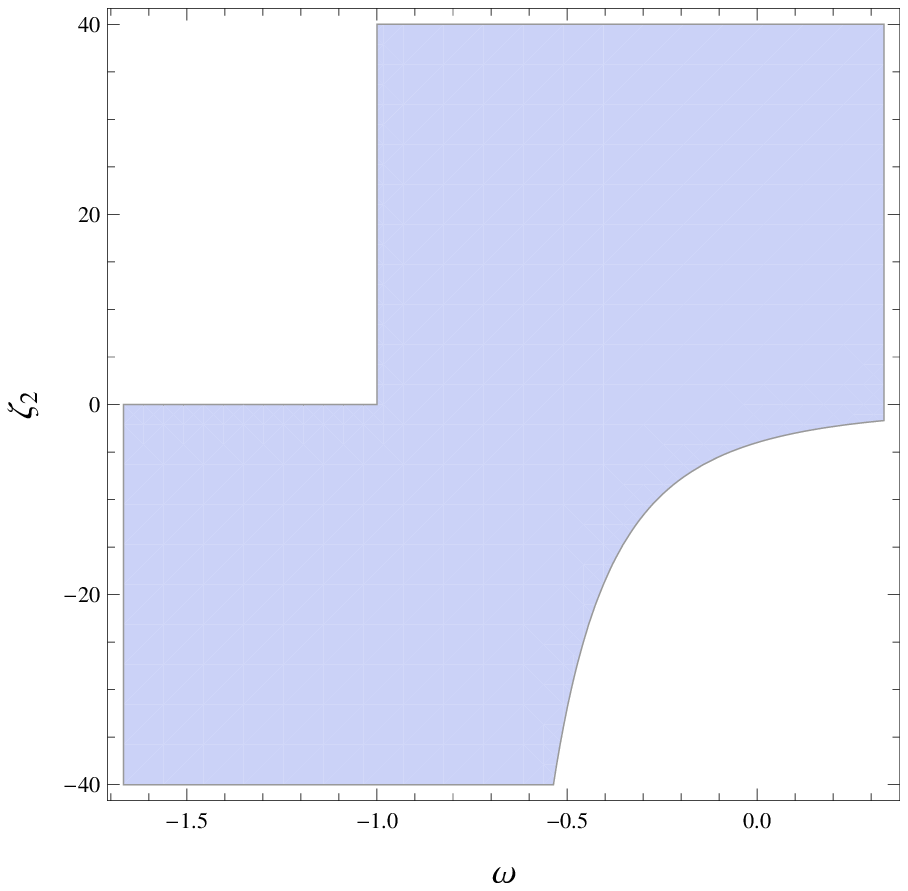,
width=0.5\linewidth}\caption{Stable regions in $(\omega,\zeta_{2})$
space for $\Upsilon_{2}^{2}$ with $\chi=1$ (left) and $\chi=-0.5$
(right).}
\end{figure}
\begin{figure}
\epsfig{file=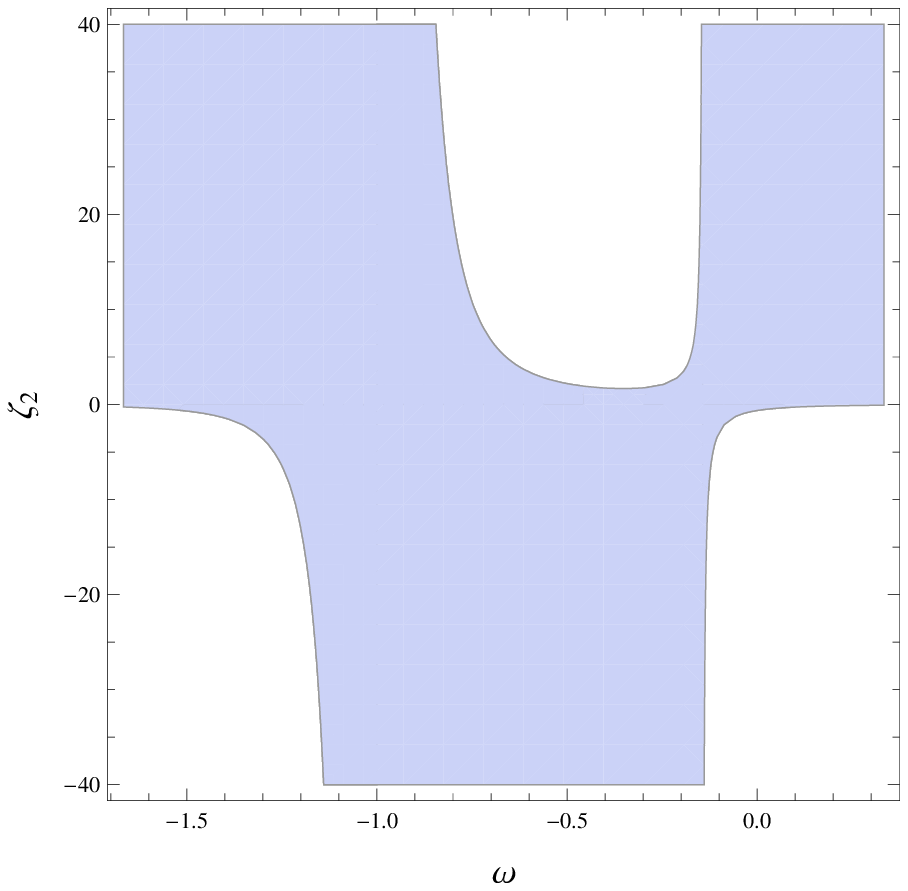, width=0.5\linewidth}\epsfig{file=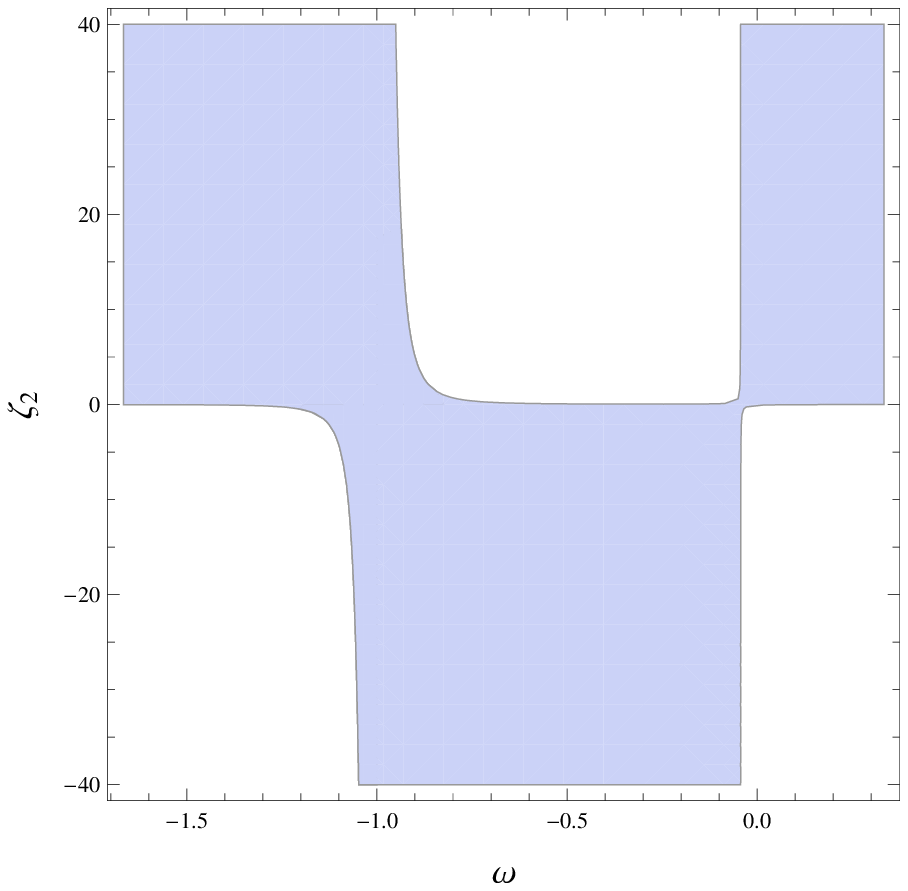,
width=0.5\linewidth}\caption{Stable regions in $(\omega,\zeta_{2})$
space for $\Upsilon_{1,2}^{2}$ with $\chi=1$ (left) and $\chi=5$
(right).}
\end{figure}

When $f_{1}(\mathcal{G}_{0})=0=\chi$, the frequencies
$\Upsilon^{2}_{1}$ recover the GR result as obtained in the previous
case while frequencies $\Upsilon^{2}_{2}$ diverge. We simplify the
expression by introducing a new parameter
$\zeta_{2}=-48\kappa^{6}\rho_{0}^{3}f_{1}''(\mathcal{G}_{0})$ which
remains positive for $f_{1}''(\mathcal{G}_{0})<0$. Figure \textbf{4}
shows stable regions against homogeneous perturbations of EU for
frequencies $\Upsilon_{1}^{2}$. It is found that when $\chi=1$ (left
panel), the stable EU exists for all values of $\omega$ with
suitable choice of $\zeta_{2}$ while less stable regions are
obtained when $\chi=5$ as shown in the right plot. In the case of
non-conserved EMT, the stability regions decrease as the value of
model parameter $\chi$ increases while no stable regions are
observed for $\chi<0$. The regions of stability in EU for
frequencies $\Upsilon^{2}_{2}$ are shown in Figure \textbf{5} for
considered values of $\chi$ while stability regions for whole system
is observed in Figure \textbf{6}.

Now, we consider the generalized model given by
\begin{equation}\label{a}
f(\mathcal{G},T)=f_{1}(\mathcal{G})+\kappa^{2}\chi T^{n},\quad
n\neq0.
\end{equation}
Following the same procedure, we obtain the following fourth-order
differential equation in perturbed $a(t)$ as follows
\begin{eqnarray}\nonumber
&&24\kappa^4\rho_{0}^{2}(1+\omega)^{2}\left[1+n\chi\rho_{0}^{n-1}(1-3\omega)
^{n-1}\right]^{2}f_{1}''(\mathcal{G}_{0})\delta
a^{(iv)}-2\delta\ddot{a}
\\\nonumber&+&\kappa^2\rho_{0}(1+\omega)\left(1+n\chi\rho_{0}^{n-1}(1-3\omega)^{n-1}\right)
\left[1+3\left(\omega-\frac{n}{2}\chi(1-3\omega)^{n}\rho_{0}^{n-1}\right)\right.
\\\nonumber&\times&\left.\left(1+n\chi(1-3\omega)^{n-1}\rho_{0}^{n-1}\left[(1+\omega)
+\frac{1}{2}(1-3\omega)+(n-1)\right.\right.\right.
\\\nonumber&\times&\left.\left.\left.(1+\omega)\right]\right)^{-1}\right]\delta
a=0,
\end{eqnarray}
whose solution provides the following four frequencies as
\begin{eqnarray}\nonumber
\Xi^{2}_{1,2}=\frac{1\pm\sqrt{1-24\kappa^6\Delta_{5}f''_{1}(\mathcal{G}_{0})}}
{24\left[\kappa^2\rho_{0}(1+\omega)\left(1+n\chi\rho_{0}^{n-1}(1-3\omega)^{n-1}
\right)\right]^{2}f_{1}''(\mathcal{G}_{0})},
\end{eqnarray}
where
\begin{eqnarray}\nonumber
\Delta_{5}&=&\rho_{0}^{3}(1+\omega)^{3}\left(1+n\chi\rho_{0}^{n-1}(1-3\omega)^{n-1}
\right)^{3}\left[1+3\left(\omega-\frac{n}{2}\chi(1-3\omega)^{n}
\right.\right.\\\nonumber&\times&\left.\left.\rho_{0}^{n-1}\right)
\left[1+n\chi(1-3\omega)^{n-1}\rho_{0}^{n-1}\left((1+\omega)+\frac{1}{2}(1-3\omega)
+(n-1)\right.\right.\right.\\\nonumber&\times&\left.\left.\left.
(1+\omega)\right)\right]^{-1}\right].
\end{eqnarray}
\begin{figure}
\epsfig{file=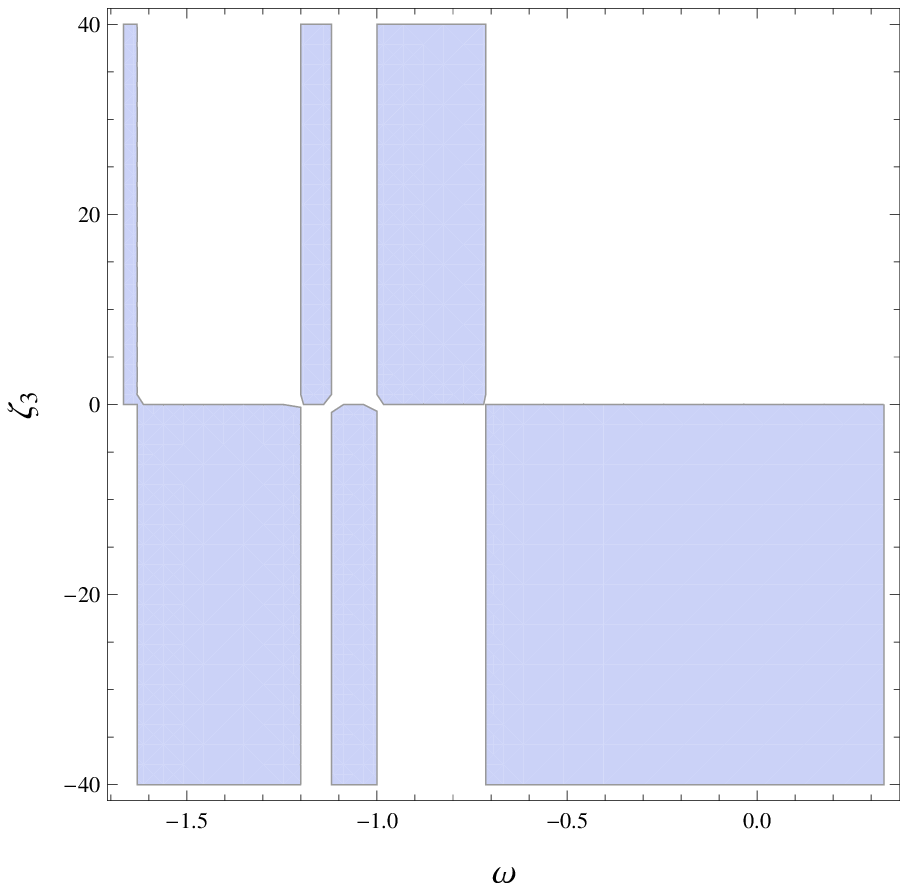, width=0.5\linewidth}\epsfig{file=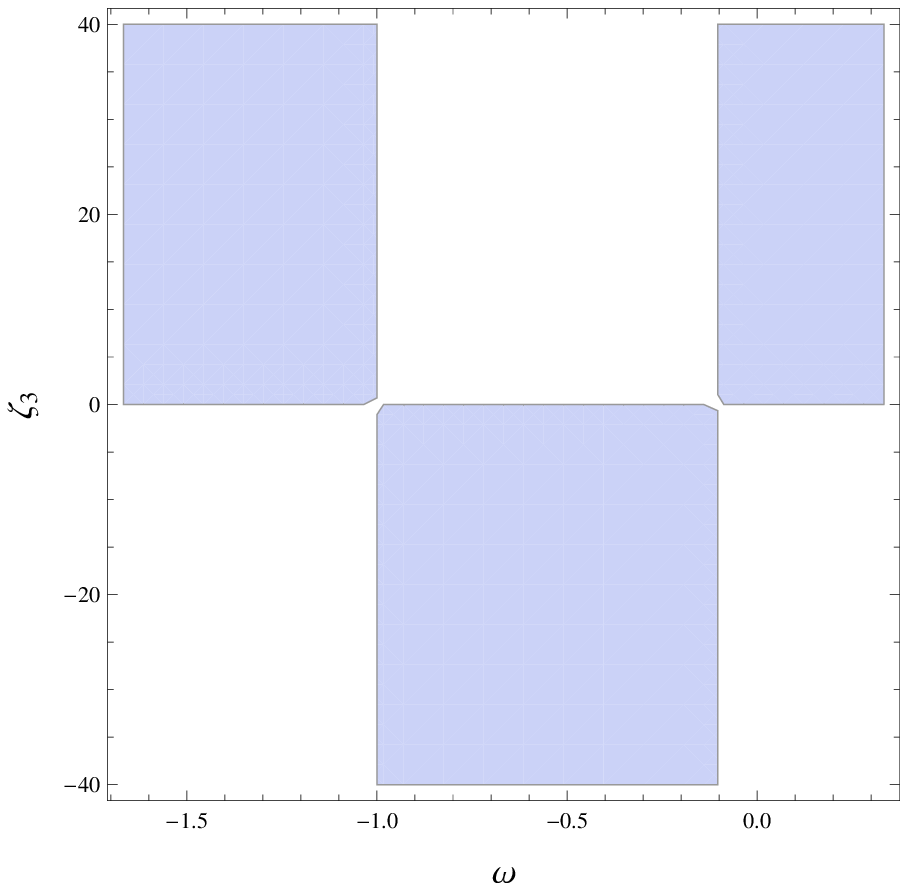,
width=0.5\linewidth}\caption{Stable regions in $(\omega,\zeta_{3})$
space for $\Xi_{2}^{2}$ with $n=-5$ (left) and $n=0.5$ (right).}
\end{figure}
\begin{figure}
\epsfig{file=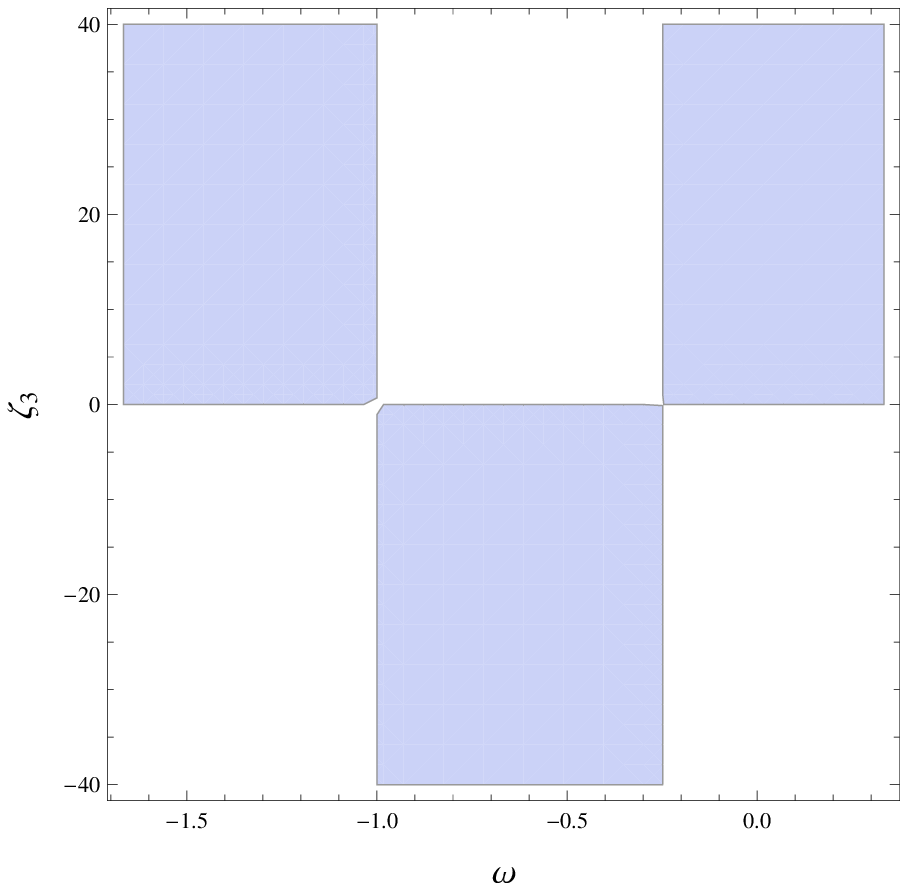, width=0.5\linewidth}\epsfig{file=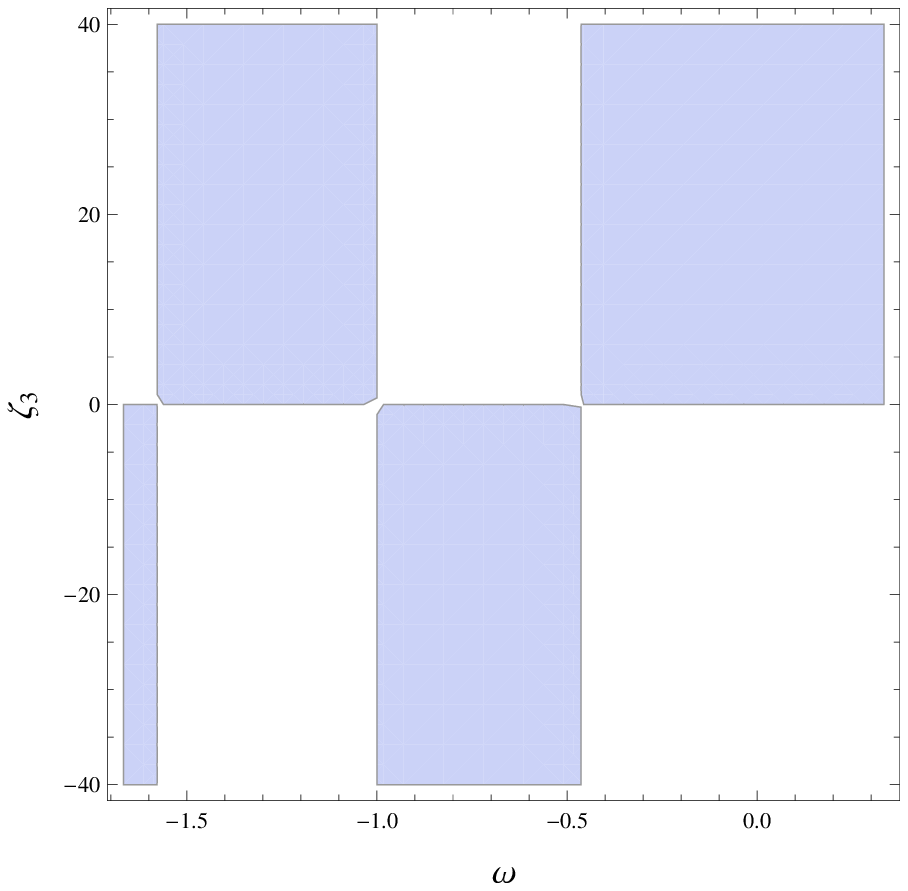,
width=0.5\linewidth}\caption{Stable regions in $(\omega,\zeta_{2})$
space for $\Xi_{2}^{2}$ with $n=2$ (left) and $n=5$ (right).}
\end{figure}
The graphical analysis of frequencies $\Xi_{2}^{2}$ are shown in
Figures \textbf{7} and \textbf{8} where we have used $\zeta_{3}=-24
\kappa^{6}f_{1}''(\mathcal{G}_{0}),~\kappa^2=1,~\rho_{0}=0.3$ and
$\chi=1$. It is found that stable regions are obtained for all the
considered values of $n$ while stable EU does not exist for the
frequencies $\Xi_{1}^{2}$. In this case, the stability region of
whole system is completely described by the frequencies
$\Xi_{2}^{2}$. It is interesting to mention here that for
$f_{1}(\mathcal{G}_{0})=0=\chi$, the frequencies $\Xi^{2}_{1}$
diverge while GR is recovered for the frequencies $\Xi^{2}_{2}$ as
in the previous case.

\section{Final Remarks}

In this paper, we have analyzed the stability issue of EU in the
context of $f(\mathcal{G},T)$ gravity which is the extension of
$f(\mathcal{G})$ gravity and is based on the ground of
matter-curvature coupling. Due to this coupling, the conservation
law does not hold as in $f(R,T)$ gravity \cite{7}. We have
considered the isotropic and homogeneous positive curvature FRW line
element with perfect fluid as matter content of the universe. The
static as well as perturbed field equations are constructed against
linear homogeneous perturbations which are parameterized by equation
of state parameter. We have formulated the fourth-order perturbed
differential equation whose solutions are analyzed for the existence
and stability of EU for specific form of
$f(\mathcal{G},T)=f_{1}(\mathcal{G})+f_{2}(T)$. For this choice, we
have discussed both the models when EMT is conserved as well not
conserved and obtained distinct results as compared to
$f(\mathcal{G})$ gravity.
\begin{itemize}
\item We have assumed that EMT is conserved in this gravity and
obtained a particular form of $f_{2}(T)$ for which the covariant
divergence of EMT becomes zero. We have analyzed the regions of
stability around EU and found that stable results are observed for a
suitable choice of integration constant $c_{1}$.
\item Two particular forms of $f_{2}(T)$ are considered for which the
covariant divergence of EMT remains non-zero and the value of energy
density in terms of scale factor is evaluated. It is found that
stable EU exists in this case for both models if the model parameter
$\chi$ is chosen appropriately.
\end{itemize}
We conclude that the stable EU universe exists against scalar
homogeneous perturbations in the background of $f(\mathcal{G},T)$
for all values of the equation of state parameter if the model
parameters are chosen suitably. Einstein universe against vector
perturbations (comoving dimensionless vorticity vector) are stable
for all equations of state on all scales since any initial vector
perturbations remain frozen. The mechanism for stability analysis of
EU against tensor perturbations (comoving dimensionless traceless
shear tensor) suggests that these fluctuations may not break the
stability of EU in the background of $f(\mathcal{G},T)$ gravity
\cite{13}. It would be interesting to investigate complete analysis
of tensor as well as inhomogeneous perturbations in this gravity
which will be helpful to explore the EU. It is worth mentioning here
that our results reduce to $f(\mathcal{G})$ gravity in the absence
of matter-curvature coupling \cite{18}.


\begin{thebibliography}{30}

\bibitem{2} Calcagni, G., Tsujikawa, S. and Sami, M.: Class. Quantum Grav.
\textbf{22}(2005)3977; De Felice, A., Hindmarsh, M. and Trodden, M.:
J. Cosmol. Astropart. Phys. \textbf{08}(2006)005.

\bibitem{3} Metsaev, R.R. and Tseytlin, A.A.: Nucl. Phys. B
\textbf{293}(1987)385; Antoniadis, I., Rizos, J. and Tamvakis, K.:
Nucl. Phys. B \textbf{415}(1994)497; Kanti, P., Rizos, J. and
Tamvakis, K.: Phys. Rev. D \textbf{59}(1999)083512; Nojiri, S. and
Odintsov, S.D.: Int. J. Geom. Meth. Mod. Phys. \textbf{4}(2007)115.

\bibitem{4} Nojiri, S. and Odintsov, S.D.: Phys. Lett. B \textbf{631}(2005)1.

\bibitem{a} Cognola, G. et al.: Phys. Rev. D \textbf{73}(2006)084007.

\bibitem{5} De Felice, A. and Tsujikawa, S.: Phys. Lett. B \textbf{675}(2009)1.

\bibitem{6} De Felice, A. and Tsujikawa, S.: Phys. Rev. D \textbf{80}(2009)063516.

\bibitem{7} Harko, T. et al.: Phys. Rev. D \textbf{84}(2011)024020.

\bibitem{8} Sharif, M. and Ikram, A.: Eur. Phys. J. C \textbf{76}(2016)640.

\bibitem{10} Eddington, A.S.: Mon. Not. R. Astron. Soc. \textbf{90}(1930)668.

\bibitem{11} Harrison, E.R.: Rev. Mod. Phys. \textbf{39}(1967)862.

\bibitem{12} Gibbons, G.W.: Nucl. Phys. B \textbf{292}(1987)784.

\bibitem{13} Barrow, J.D. et al.: Class. Quantum Grav. \textbf{20}(2003)L155.

\bibitem{b} Ellis, G.F.R. and Maartens, R.: Class. Quantum Grav.
\textbf{21}(2004)223; Ellis, G.F.R., Murugan, J. and Tsagas, C.G.:
Class. Quantum Grav. \textbf{21}(2004)233.

\bibitem{14} Gergely, L.\'{A}. and Maartens, R.: Class. Quantum. Grav. \textbf{19}(2002)213;
Gruppuso, A. Roessl, E. and Shaposhnikov, M.: J. High Energy Phys.
\textbf{08}(2004)011; B\"{o}hmer, C.G.:  Class. Quantum. Grav.
\textbf{21}(2004)1119; Mulryne, D.J. et al.: Phys. Rev. D
\textbf{71}(2005)123512; Atazadeh, K. and Darabi, F.: Phys. Lett. B
\textbf{744}(2015)363; Zhang, K. et al.: Phys. Lett. B
\textbf{758}(2016)37.

\bibitem{15} B\"{o}hmer, C.G., Hollenstein, L. and Lobo, F.S.N.: Phys.
Rev. D \textbf{76}(2007)084005.

\bibitem{16} Goswami, R., Goheer, N. and Dunsby, P.K.S.: Phys.
Rev. D \textbf{78}(2008)044011.

\bibitem{17} Goheer, N. Goswami, R. and Dunsby, P.K.S.: Class. Quantum Grav.
\textbf{26}(2009)105003.

\bibitem{18} B\"{o}hmer, C.G. and Lobo, F.S.N.: Phys. Rev. D
\textbf{79}(2009)067504.

\bibitem{c} B\"{o}hmer, C.G., Lobo, F.S.N. and Tamanini, N.: Phys. Rev. D \textbf{88}(2013)104019.

\bibitem{cc} Li, J.T., Lee, C.C. and Geng, C.Q.: Eur. Phys. J. C
\textbf{73}(2013)2315.

\bibitem{d} Huang, H., Wu, P. and Yu, H.: Phys. Rev. D \textbf{89}(2014)103521.

\bibitem{e} Huang, H., Wu, P. and Yu, H.: Phys. Rev. D \textbf{91}(2015)023507.

\bibitem{f} B\"{o}hmer, C.G., Tamanini, N. and Wright, M.: Phys. Rev. D \textbf{92}(2015)124067.

\bibitem{g} Darabi, F., Heydarzade, Y. and Hajkarim, F.: Can. J. Phys. \textbf{93}(2015)1566.

\bibitem{19} Shabani, H. and Ziaie, A.H.: arXiv:1606.07959.

\bibitem{20} Landau, L.D. and Lifshitz, E.M.: \emph{The Classical Theory of Fields}
(Pergamon Press, 1971).

\bibitem{h2} Ade, P.A.R. et al.: Astron. Astrophys. \textbf{594}(2016)A13.

\bibitem{21} Li, B., Barrow, J.D. and Mota, D.F.: Phys. Rev. D \textbf{76}(2007)044027.

\end{thebibliography}
\end{document}